# Twist-Angle Engineering of Moiré Potentials for High-Performance Ionics in Bilayer Graphene

[1]Gen Fukuzawa, [1,†]Yebin Lee, and [1,*]Teruyasu Mizoguchi

[1]Institute of Industrial Science, The University of Tokyo

[*]Corresponding author: teru@iis.u-tokyo.ac.jp

[†]Present address: National Institute for Materials Science

## Abstract

Controlling ion transport is a fundamental challenge for advanced energy storage. Bilayer graphene offers a unique platform for modulating ion diffusion via twist-angle-dependent moiré superlattices, yet conventional stacking configurations face an inherent trade-off: AA stacking provides stable Li intercalation but high diffusion barriers, while AB stacking enables fast diffusion but poor intercalation stability. Twisted bilayer graphene (tBLG) offers potential to overcome this limitation, yet systematic understanding across different twist angles remains limited. Here, we investigate Li intercalation in tBLG using first-principles density functional theory, evaluating intercalation energies and diffusion barriers across multiple twist angles through potential energy surface (PES) mapping. The Σ37 structure (9.43°) simultaneously achieves the most favorable intercalation energy (−2.39 eV) and the lowest diffusion barrier (0.14 eV) among all structures examined, resolving the conventional stacking trade-off. Furthermore, using the Smooth Overlap of Atomic Positions (SOAP) descriptor, we demonstrate that the PES is governed by local atomic environments and that a model trained on limited structures predicts the PES of untested configurations with high accuracy. This transferability enables efficient screening without exhaustive first-principles calculations, establishing a systematic framework for twist-angle engineering of ion transport in two-dimensional layered materials.

## 1. Introduction

Ionic conductivity in solid materials is a crucial performance parameter for electrochemical energy storage devices, including lithium-ion batteries. In anode materials particularly, both lithium ions and electrons must move efficiently, and their transport properties are governed by the crystal structure of the material[1]. Since battery power density, charge-discharge rate, and cycle life all strongly depend on ion mobility, fundamental understanding of ion conduction mechanisms is essential for developing high-performance battery materials[2–4]. In recent years, with the growing adoption of electric vehicles and large-scale energy storage systems, there is strong demand for battery materials with higher energy density and rapid charging capabilities[3,5].

In crystalline materials, lithium ions move through interstitial sites via a hopping mechanism, making the geometry of diffusion pathways and the height of activation energy barriers important factors determining ionic conductivity[6,7]. The diffusion process is described as an activated process in which lithium ions move from initial stable sites to adjacent sites via transition states[8]. The dimensionality of diffusion pathways is also important. For example, layered oxides such as LiCoO2 have two-dimensional diffusion pathways within the layers[9], while spinel structures such as LiMn2O4 allow three-dimensional diffusion[10]. In general, crystal structures with wider diffusion pathways and lower activation energy barriers exhibit higher ionic conductivity[7,11]. Additionally, the vacancy size and coordination environment in crystal structures significantly affect lithium ion intercalation energy and diffusion barriers[12]. In particular, when the bottleneck size of diffusion pathways is small, lithium ions experience strong repulsive interactions with surrounding atoms, leading to a significant increase in diffusion barriers[13]. Therefore, precise control of crystal structure may enable dramatic improvements in ion conduction properties[11,12].

In recent years, two-dimensional materials have attracted attention for their unique electronic and structural properties[14]. In particular, graphene is expected as a next-generation battery material due to its high electrical conductivity, excellent mechanical strength, and large specific surface area[15,16].

In monolayer graphene, carbon atoms form a planar honeycomb structure through sp2 hybridization, and the π-electron system extends in the in-plane direction, achieving extremely high electron mobility[17]. The theoretical capacity of graphene as a lithium-ion battery anode significantly exceeds that of graphite when considering adsorption on both sides. Furthermore, in bilayer graphene (BLG), the stacking of two graphene layers changes the electronic states through interlayer interactions, exhibiting different properties from monolayer graphene[18]. In particular, it is known that the band structure and Fermi velocity change significantly depending on the stacking configuration (AA stacking, AB stacking, etc.)[19]. AB-stacked BLG shows parabolic dispersion in the low-energy region[19,20], while AA-stacked BLG is known to show linear dispersion similar to monolayer graphene[19].

Since the discovery of superconductivity near the "magic angle" (~1.1°) in twisted bilayer graphene (tBLG) by Cao et al.[21], moiré structure systems in which properties can be controlled by twist angle have attracted great attention[22,23]. In tBLG, twisting two graphene layers at a specific angle creates periodic moiré patterns[24]. This moiré structure has a dramatic effect on electronic states, and at specific twist angles, flat bands are formed, making electron correlation effects prominent[21,25]. Additionally, the interlayer distance and local stacking configuration change with twist angle, which may affect not only electronic properties but also mechanical properties and ion intercalation characteristics[26,27].

Many studies have reported on the application of graphene-based materials to lithium-ion battery anodes[15,16,28]. For monolayer graphene, lithium adsorption energy and diffusion barriers have been investigated in detail[29,30]. For multilayer graphene and graphite, lithium intercalation between layers is the main reaction, and the stage structure formation mechanism is becoming clear[31–33]. However, most of these studies focus on monolayer graphene or multilayer graphene with uniform stacking structures, and there is a lack of systematic understanding of the lithium intercalation properties of tBLG. Recent theoretical studies have reported that lithium preferentially clusters in AA regions in tBLG[34], but the effects of twist angle and stacking structure on lithium intercalation energy and diffusion barriers have not been systematically elucidated. In particular, how differences in local stacking configurations formed by moiré structures (AA regions, AB regions, saddle point regions) affect the stable sites and diffusion pathways of lithium ions[26,34] is an important unresolved question. Recent experimental studies have confirmed multiple different in-plane staging phases in bilayer graphene and shown that diffusion barriers in AB stacking are significantly lower than in AA stacking[35], but systematic studies of tBLG with arbitrary twist angles have rarely been conducted.

Therefore, in this study, we utilize tBLG as a model platform to systematically investigate how moiré-induced local atomic environments govern ion intercalation, namely "Moiré-ionics", employing first-principles calculations across multiple twist angles. Specifically, for AA stacking, AB stacking, and multiple tBLG structures, we quantitatively evaluated both diffusion barriers and intercalation energies through potential energy surface (PES) calculations on dense grids. Furthermore, we analyzed the correlation between local atomic configurations and energy characteristics, and verified whether the correlation obtained for one structure can be applied to structures with other twist angles. The results of this study are expected to provide design guidelines for high-performance lithium-ion battery anode materials using tBLG.

## 2. Results and Discussion

We investigated AA stacking, AB (Bernal) stacking, and five tBLG structures characterized by their coincidence site lattice (CSL) Σ values: Σ7 (21.79°), Σ13 (32.20°), Σ19 (13.17°), Σ31 (17.90°), and Σ37 (9.43°). **Figure 1** shows the atomic configurations of all structures investigated. Structural parameters including interlayer distances and binding energies for all structures are summarized in the Supporting Information (Table S1).

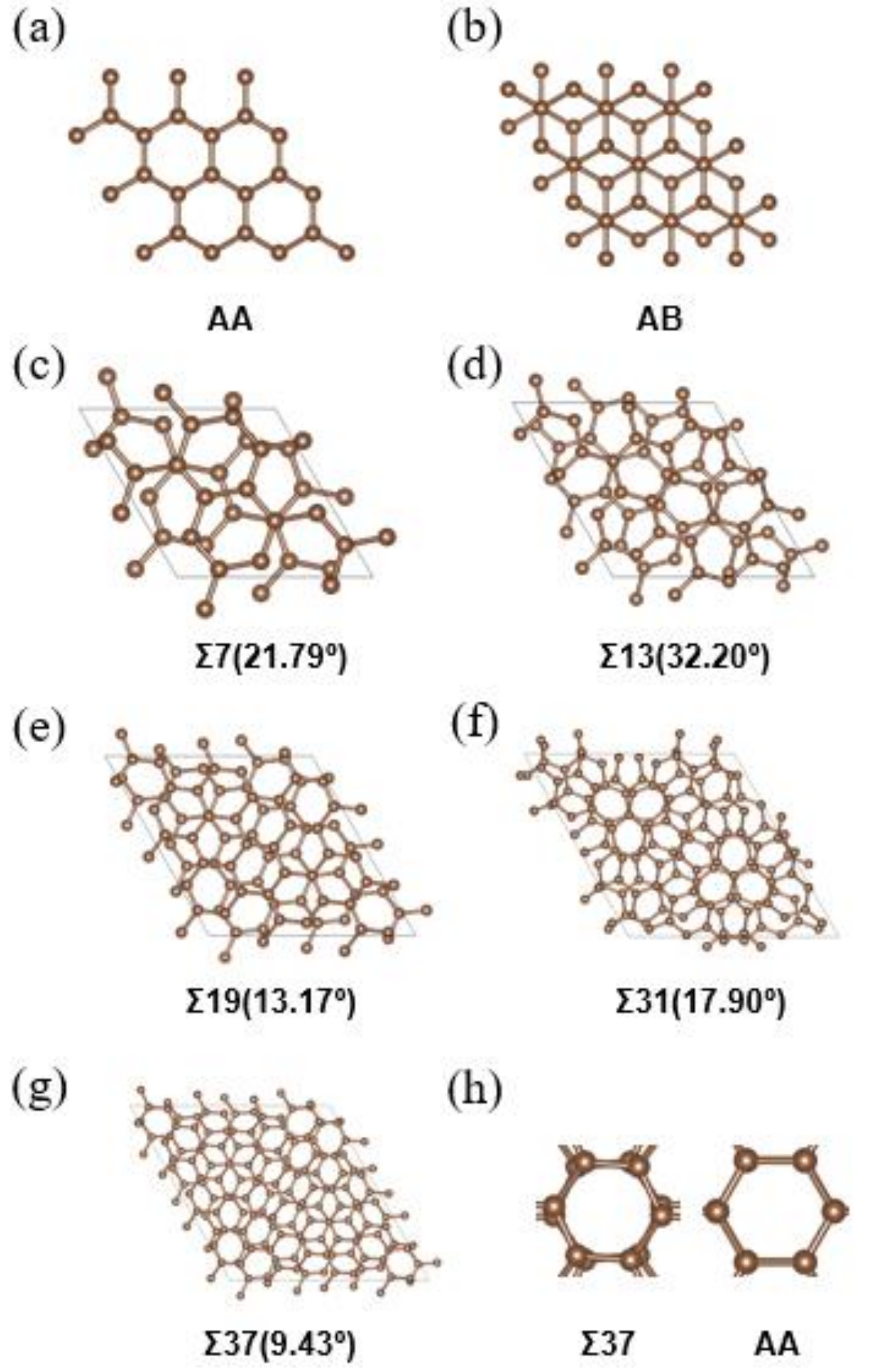

**Figure 1.** Atomic configurations of the bilayer graphene structures investigated in this study: (a) AA stacking, (b) AB stacking, (c) Σ7, (d) Σ13, (e) Σ19, (f) Σ31, and (g) Σ37. Twist angles are indicated in parentheses. (h) Local atomic environment at the twist rotation center of the Σ37 structure (left) compared with the hollow site of AA stacking (right), viewed from above.

### 2.1 Potential Energy Surface and Li Intercalation Properties

### 2.1.1 Potential Energy Surface

The potential energy surface (PES) for Li intercalation was computed for AA stacking, AB stacking, and the five tBLG structures (Σ7–Σ37) by systematically placing a single Li atom at grid points within the interlayer space and relaxing the atomic positions along the z-axis (see Methodology for details). The PES represents the relative energy at each grid point, referenced to the most stable intercalation site within each structure.

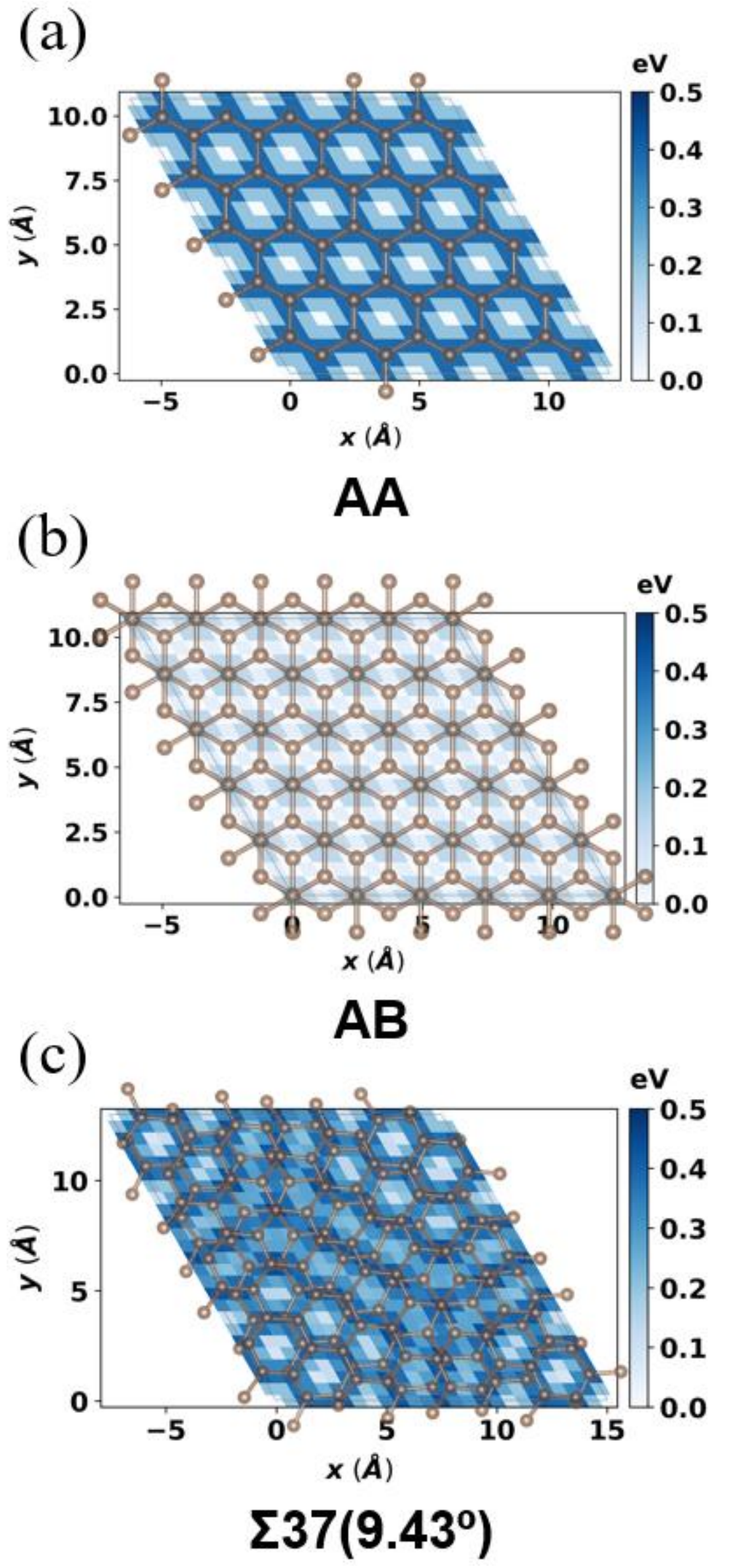

**Figure 2.** Potential energy surface maps for Li intercalation in (a) AA stacking, (b) AB stacking, and (c) the Σ37 structure. The color scale represents the relative energy at each grid point, referenced to the most stable intercalation site within each structure.

**Figure 2** shows representative PES maps for AA stacking, AB stacking, and the Σ37 structure. The observed PES patterns can be understood from the site-dependent adsorption energies of Li on monolayer graphene. On a single graphene sheet, the hollow site (center of the hexagonal ring) is the most stable adsorption site for Li (−1.25 eV), while the bridge (−0.96 eV) and on-top (−0.95 eV) sites are significantly less favorable and nearly degenerate with each other (Figure S1). The primary energetic distinction is therefore between the hollow site and the two non-hollow sites, with an energy difference of approximately 0.3 eV.

In AA stacking, the two graphene layers are perfectly aligned, so the hollow, bridge, and on-top positions of both layers coincide. A Li atom at the interlayer hollow site benefits from the favorable hollow-site environment of both layers simultaneously, making it the most stable position. Conversely, the on-top site is unfavorable with respect to both layers, resulting in the highest energy. This reinforcement produces a PES with large energy variation and the pronounced periodic pattern seen in Figure 2a.

In AB stacking, the lateral offset between the two layers alters the site correspondence. While positions that are on-top with respect to both layers still exist, positions that are simultaneously hollow for both layers do not—a site that is hollow in one layer corresponds to an on-top position in the other. As a result, the most favorable environment present in AA stacking (double hollow) is absent, and the overall energy variation is reduced compared to AA stacking (maximum approximately 0.22 eV). This smoothing of the energy landscape creates continuous low-energy pathways that enable Li diffusion with a remarkably low barrier, as discussed in Section 2.1.3.

The tBLG structures exhibit PES patterns that reflect their moiré superlattice periodicity. As shown for the Σ37 structure (Figure 2c), the PES contains spatially varying regions of high and low energy that correspond to the local stacking environment within the moiré pattern. Regions where the local stacking resembles AA-like configurations exhibit higher energy variation, while regions with AB-like local stacking show flatter energy landscapes. The complete PES maps for all structures, including selected diffusion pathways, are provided in the Supporting Information (Figure S2).

### 2.1.2 Intercalation Energy

The intercalation energy at the most stable site of each structure is shown in **Figure 3a**. The intercalation energies reported here were computed using supercells large enough to ensure convergence with respect to cell size, so that the observed variation across structures reflects differences in local atomic environments rather than finite-size artifacts. The intercalation energy varies substantially across structures, ranging from −1.87 eV (Σ13) to −2.39 eV (Σ37).

Among the reference structures, AA stacking shows a moderately favorable intercalation energy of −2.19 eV, while AB stacking is less stable at −1.97 eV. Within the tBLG structures, Σ37 exhibits the most favorable intercalation energy (−2.39 eV), followed by Σ31 (−2.36 eV). In contrast, Σ13 shows the least favorable value (−1.87 eV), which is even less stable than AB stacking.

Notably, the intercalation energy does not exhibit a simple monotonic relationship with either the twist angle or the initial interlayer distance. For example, AA stacking, which has the largest interlayer distance (3.52 Å; Table S1), does not yield the most favorable intercalation energy, and the Σ37 structure, with a relatively compact interlayer spacing (3.41 Å), achieves the most stable value. This result indicates that intercalation stability is not governed solely by long-range structural parameters, but rather depends on the details of the local atomic arrangement surrounding the intercalation site—a point that we examine quantitatively in Section 2.2.

### 2.1.3 Diffusion Barrier

To evaluate the kinetic properties of Li transport, diffusion barriers were extracted from the PES for each structure. Diffusion pathways were identified by tracing low-energy routes that traverse the entire periodic cell, representing the minimum energy path for sustained Li diffusion in a given direction (see Supporting Information, Figure S2 for the selected pathways and energy profiles). The diffusion barrier is defined as the energy difference between the maximum and minimum along this pathway.

Figure 3b summarizes the diffusion barriers for all structures. The barriers span a wide range, from 0.04 eV (AB stacking) to 0.39 eV (AA stacking). In AA stacking, the two layers are perfectly aligned, so the hollow/on-top energy difference of each layer (~0.3 eV on monolayer graphene; Figure S1) is reinforced across both layers, producing deep energy minima separated by high barriers (0.39 eV) and severely restricting Li diffusion. In AB stacking, the absence of double-hollow sites eliminates this reinforcement, and the energy landscape is instead governed by the near-degenerate bridge and on-top sites (~0.01 eV difference on monolayer graphene), resulting in an exceptionally flat and continuous low-energy pathway with a barrier of only 0.04 eV.

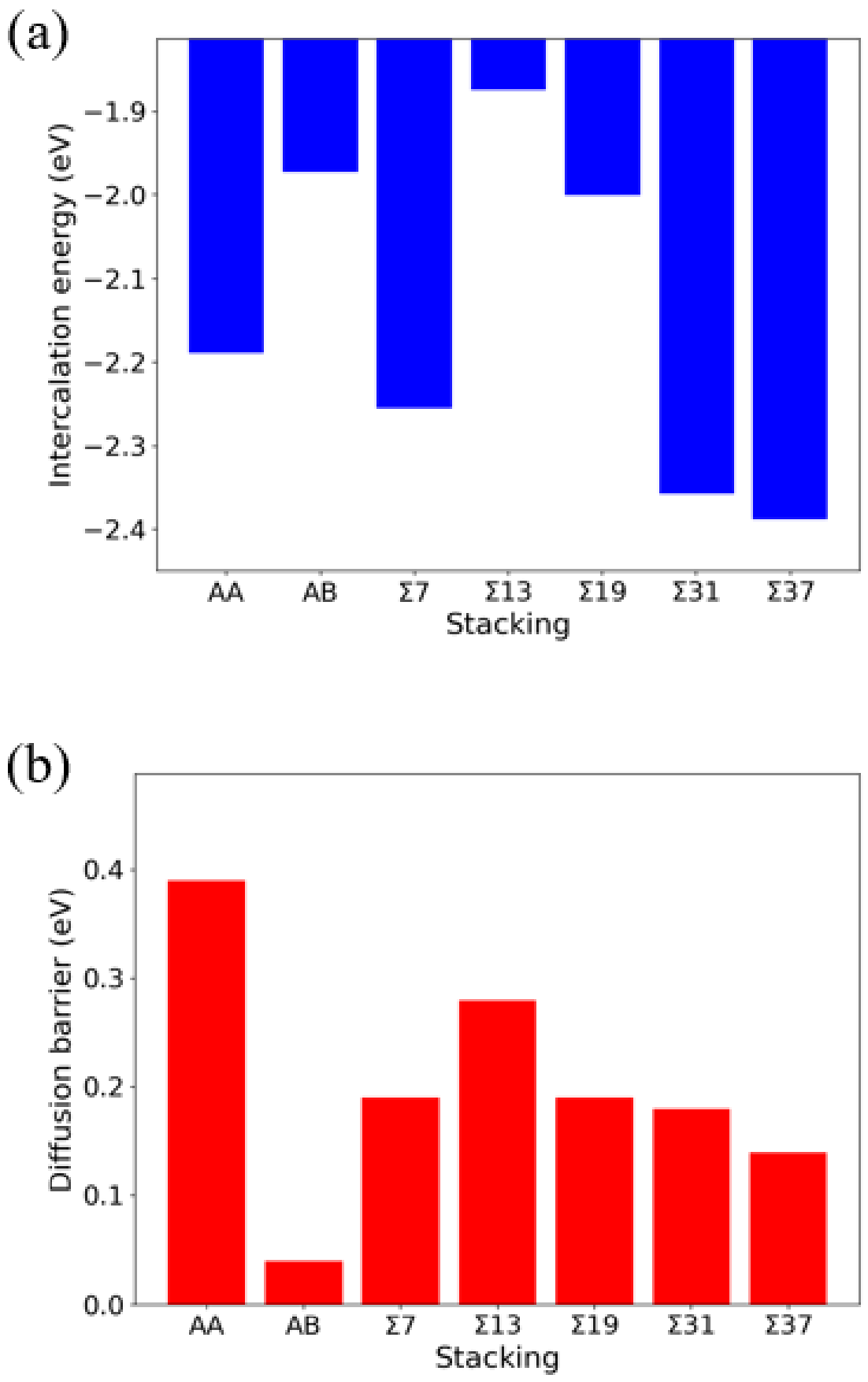

**Figure 3.** (a) Intercalation energy at the most stable site and (b) diffusion barrier for each structure. More negative intercalation energies indicate greater thermodynamic stability. Lower diffusion barriers indicate more facile Li transport.

The tBLG structures show intermediate diffusion barriers ranging from 0.14 to 0.28 eV. Among these, Σ37 exhibits the lowest barrier (0.14 eV), followed by Σ31 (0.18 eV), Σ7 (0.19 eV), and Σ19 (0.19 eV). Σ13 shows the highest barrier among the tBLG structures (0.28 eV). The variation in diffusion barriers across different tBLG structures indicates that the twist angle significantly influences the kinetic properties of Li transport, although the relationship is not a simple monotonic function of the twist angle or the moiré period.

Although AB stacking achieves the lowest diffusion barrier among all structures examined, its intercalation energy (−1.97 eV) is comparatively unfavorable, indicating that fast diffusion alone does not make it an optimal electrode material. This trade-off between kinetic and thermodynamic properties is discussed in detail in the following section.

### 2.1.4 Thermodynamic–Kinetic Trade-off

For practical battery electrode applications, both thermodynamic stability (favorable intercalation energy for high energy density) and kinetic accessibility (low diffusion barrier for fast charge/discharge rates) are essential. The results presented in Sections 2.1.2 and 2.1.3 reveal a notable trade-off between these two properties in the conventional stacking configurations.

AA stacking provides a relatively stable intercalation energy (−2.19 eV) but suffers from a high diffusion barrier (0.39 eV), which would severely limit the charge/discharge rate. AB stacking, conversely, offers an extremely low diffusion barrier (0.04 eV) but exhibits comparatively unfavorable intercalation stability (−1.97 eV). Neither conventional stacking configuration simultaneously satisfies both requirements.

The Σ37 structure resolves this trade-off. With the most favorable intercalation energy among all structures examined (−2.39 eV) and the lowest diffusion barrier among the tBLG structures (0.14 eV), Σ37 simultaneously achieves both thermodynamic stability and kinetic accessibility. Compared to AA stacking, Σ37 improves the intercalation energy by approximately 0.2 eV while reducing the diffusion barrier by 0.25 eV. These results identify Σ37 as the most promising candidate among the structures investigated for Li-ion battery electrode applications, and demonstrate that twist angle control provides an effective means of optimizing the Li intercalation properties of bilayer graphene.

### 2.2 Local Atomic Environment and PES Prediction

As discussed in Section 2.1.2, the intercalation energy does not correlate simply with long-range structural parameters such as twist angle or interlayer distance. This motivates an examination of whether the PES is instead governed by the local atomic environment surrounding each Li site. In this section, we first demonstrate this relationship qualitatively using a simple geometric descriptor, then show that a high-dimensional descriptor achieves quantitative prediction accuracy both within and across structures.

#### 2.2.1 Qualitative Correlation with Local Interatomic Distances

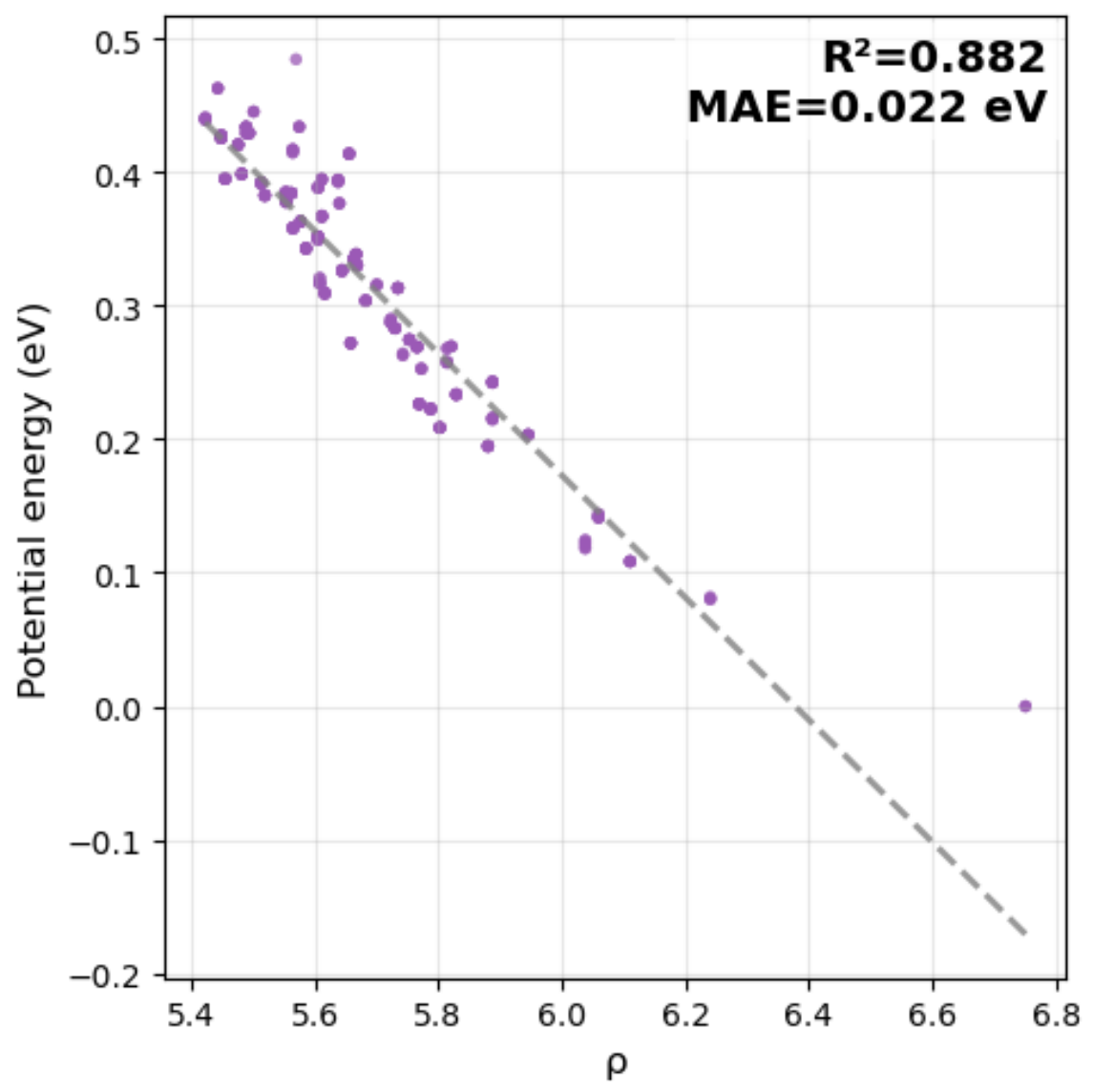

**Figure 4.** Correlation between the distance sum descriptor ρ (sum of the three nearest Li–C distances) and potential energy for the Σ37 structure. The dashed line indicates a linear fit.

To test whether the local atomic environment influences the PES, we defined a simple scalar descriptor ρ as the sum of the distances from the Li atom to its three nearest C atoms:

$$\rho = \Sigma\, r_i (i = 1,2,3)$$

where $r_i$ is the distance from the Li position to the i-th nearest C atom. **Figure 4** shows the relationship between ρ and the potential energy for the Σ37 structure. A clear negative correlation is observed: sites where the nearest C atoms are closer (smaller ρ) tend to have higher potential energy, while sites with more distant neighbors (larger ρ) are more stable. Linear regression within each structure yields $R^2$ values of 0.85–0.89 across all tBLG structures.

This result provides direct evidence that the PES is governed primarily by the local atomic arrangement around the Li site, rather than by long-range structural parameters. However, the overall scatter ($R^2 < 0.9$) indicates that a scalar distance descriptor does not fully capture the complexity of the local environment. The deviation is particularly pronounced at a site near the twist rotation center of the moiré pattern (Figure 1h), which appears as the isolated outlier in the lower-right region of Figure 4. This site has an exceptionally large ρ (~6.7) because the local environment resembles the hollow site of AA stacking with neighboring C atoms distributed at relatively large and uniform distances. Although this site corresponds to the most stable intercalation position (potential energy=0eV by definition), the linear model, extrapolating the trend of "larger ρ is more stable", predicts an unphysically negative value of approximately -0.2 eV, significantly overestimating the stability.

Beyond this outlier, residual scatter persists throughout the entire ρ range ($R^2$ = 0.882, MAE = 0.022 eV), indicating that the scalar distance descriptor does not fully capture the complexity of the local environment. This is physically reasonable: two sites can share the same sum of nearest-neighbor distances yet differ substantially in the angular arrangement of surrounding C atoms, leading to different intercalation energies. These limitations motivate the use of a more comprehensive descriptor that encodes the full three-dimensional arrangement of neighboring atoms.

#### 2.2.2 Quantitative Prediction Using the SOAP Descriptor

To capture the full complexity of the local atomic environment, we employed the Smooth Overlap of Atomic Positions (SOAP) descriptor [36], which encodes the three-dimensional arrangement of neighboring atoms as a rotationally invariant power spectrum of the neighbor density. SOAP descriptors were computed using the DScribe package[37] with a cutoff radius of 3Å and expansion parameters $n_{max} = l_{max} = 3$, yielding a descriptor vector of 84 dimensions for each Li position. Ridge regression (L2-regularized linear regression) was used to map the SOAP descriptor to the potential energy. For all structures examined (Σ7, Σ13, Σ19, Σ31, Σ37), within-structure prediction achieved $R^2 > 0.99$ (**Figure 5**), a substantial improvement over the simple distance descriptor ($R^2 \approx 0.85$–$0.89$). This confirms that the PES of each tBLG structure is essentially determined by the local atomic environment, and that the multi-dimensional SOAP representation captures the information that the scalar distance descriptor misses.

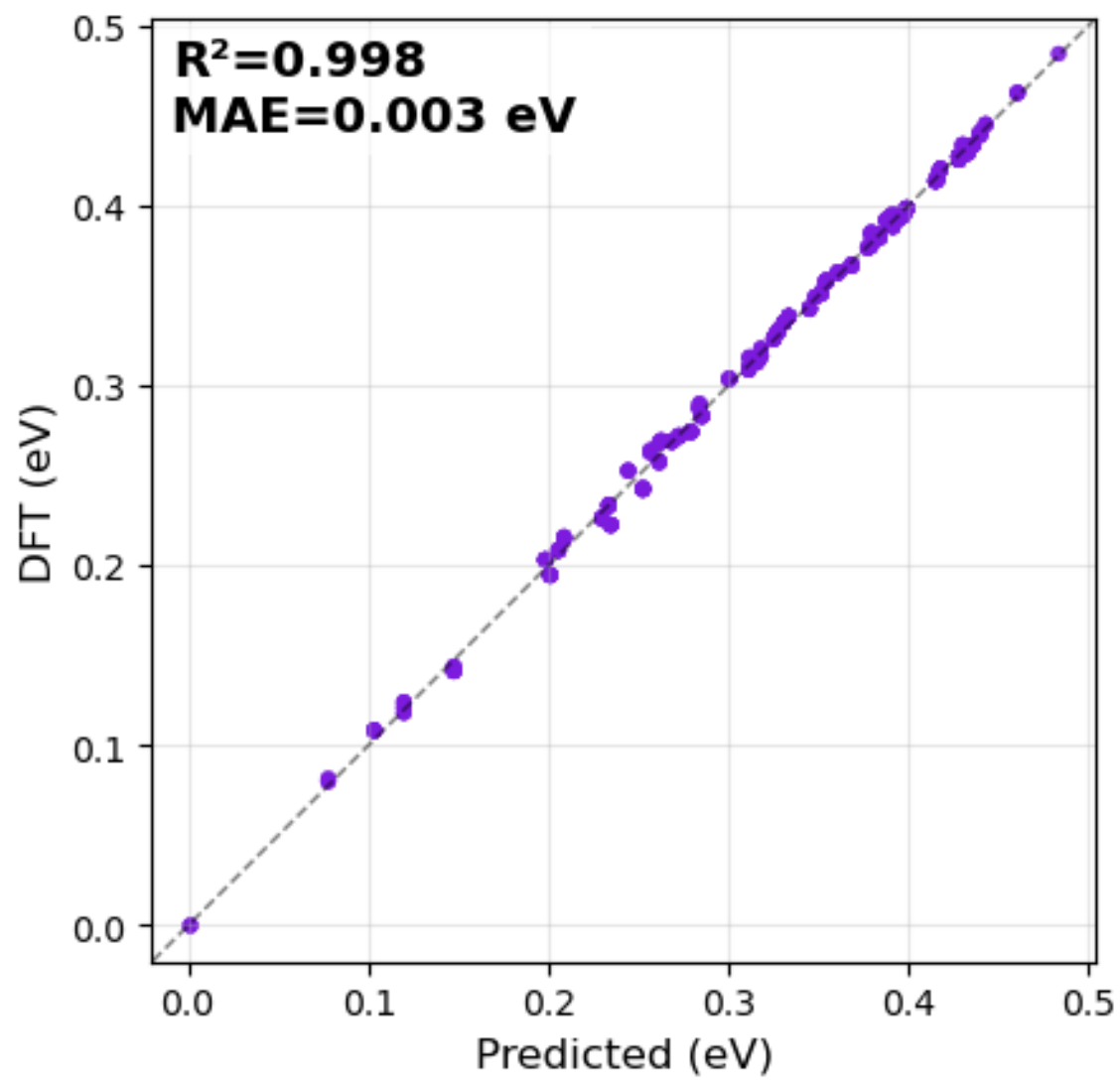

**Figure 5.** Parity plot of SOAP-predicted versus DFT-calculated potential energies for within-structure prediction of the Σ37 structure. The dashed line indicates perfect agreement.

### 2.2.3 Cross-Structure Transferability

A key question is whether the SOAP-based model trained on one set of tBLG structures can accurately predict the PES of a different structure with a distinct twist angle. If such transferability holds, the PES of an untested tBLG configuration could be predicted from existing data without performing additional DFT calculations, substantially reducing the computational effort required to screen candidate structures for optimal Li intercalation properties.

**Figure 6** shows the predicted versus DFT-calculated PES for Σ31, using a model trained on Σ13 data. The prediction achieves $R^2 = 0.994$ and MAE = 0.006 eV, demonstrating that PES data from a single structure can quantitatively predict the energy landscape of a structurally distinct configuration with near-DFT accuracy.

**Table 1** summarizes the cross-structure prediction accuracy for various training–test combinations. Two notable trends emerge. First, all cross-structure predictions achieve $R^2 > 0.9$, indicating that the relationship between local atomic environment and potential energy captured by the SOAP descriptor is transferable across different twist angles. Second, certain combinations of structures yield particularly high prediction accuracy ($R^2 > 0.99$), while others show slightly reduced performance ($R^2 \approx 0.92$). This variation suggests that the distribution of local environments sampled by different tBLG structures is not uniform; structures with similar distributions of local stacking configurations exhibit higher mutual predictability.

### 2.2.4 Physical Interpretation

The progression from the simple distance descriptor ($R^2 \approx 0.85$–$0.89$) to the SOAP representation ($R^2 > 0.99$) reveals the nature of the local environment's influence on the PES. The nearest-neighbor distances alone account for the majority of the variance, confirming that the PES is primarily determined by the immediate coordination environment of the Li atom. The improvement achieved by SOAP indicates that angular and higher-order structural features beyond a simple distance measure also contribute to determining the intercalation energy.

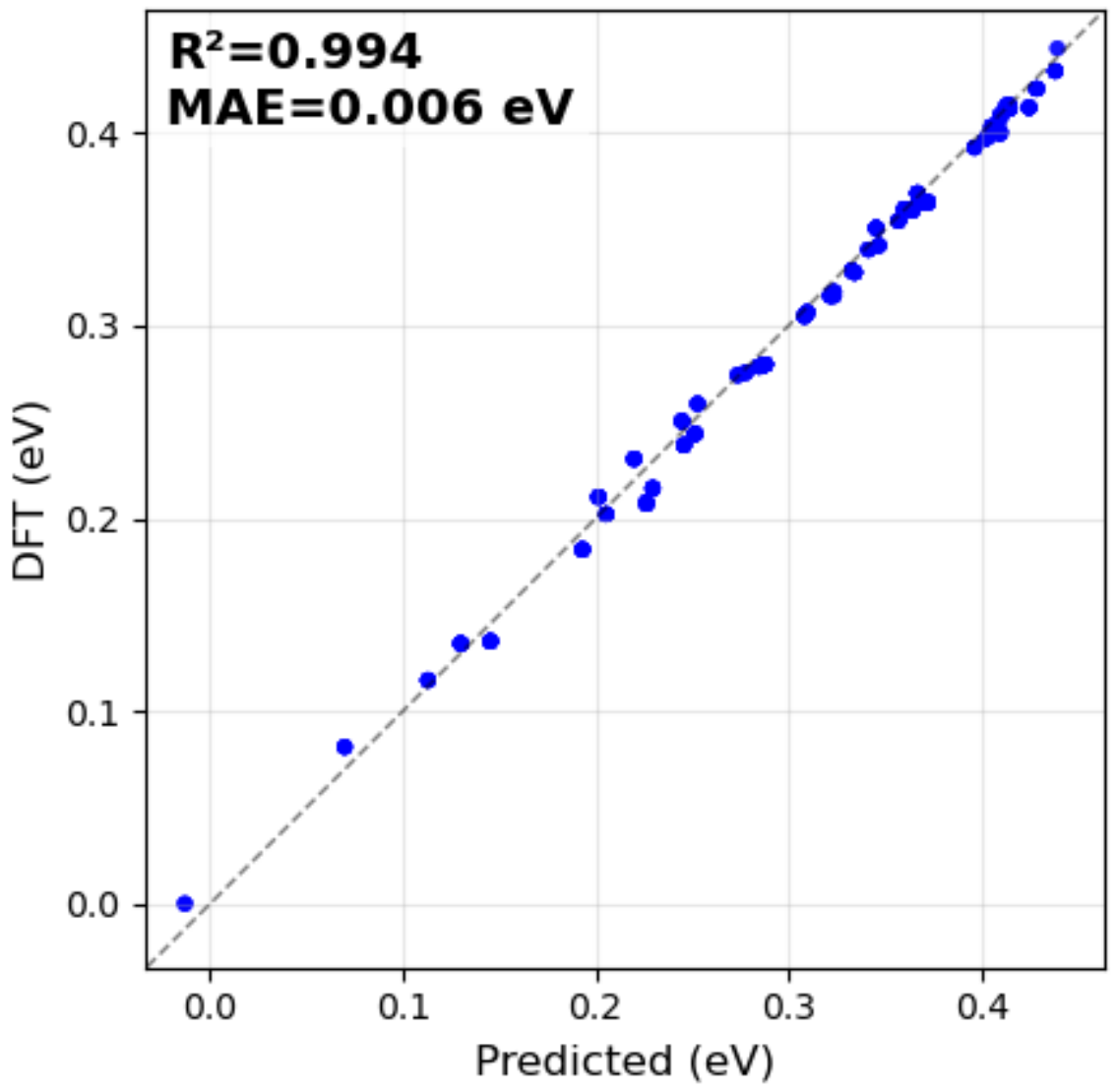

**Figure 6.** Parity plot of predicted versus DFT-calculated potential energies for the Σ31 structure, using a SOAP + Ridge regression model trained on Σ13 data. The dashed line indicates perfect agreement.

Table 1. Cross-structure prediction accuracy for representative training–test combinations using the SOAP + Ridge regression model.

| Training set | Test set | $R^2$ | MAE (eV) |
|---|---|---|---|
| Σ13 | Σ31 | 0.994 | 0.006 |
| Σ19 | Σ37 | 0.991 | 0.007 |
| Σ13 | Σ37 | 0.936 | 0.018 |
| Σ19 | Σ31 | 0.924 | 0.025 |

Notably, the SOAP descriptor used here has only 84 dimensions with a compact cutoff radius of 3 Å—comparable to the nearest Li–C interatomic distance—yet achieves near-perfect within-structure prediction. Extending the cutoff radius to larger values does not improve the accuracy, confirming that the relevant structural information is concentrated in the immediate vicinity of the Li site. This is consistent with the strong performance of the three-nearest-neighbor distance descriptor and establishes that the PES in tBLG is governed by the immediate local atomic environment, with negligible contribution from longer-range structural features.

These results establish that the PES of tBLG structures is governed by local atomic environments in a manner that is transferable across different twist angles. This transferability validates the use of SOAP-based models as an efficient screening tool: once PES data have been obtained for even a single tBLG structure, the model can predict the PES of additional twist angle configurations without requiring further DFT calculations, enabling rapid identification of structures with favorable Li intercalation properties.

### 2.3 Structural Design Guidelines and Outlook

The findings of this study provide guidelines for the design of tBLG-based electrode materials and suggest directions for future research.

The identification of Σ37 (9.43°) as the optimal structure among those investigated demonstrates that twist angle control offers a viable route to simultaneously optimize thermodynamic stability and kinetic accessibility for Li intercalation. This result highlights that the performance of bilayer graphene electrodes is not limited to the properties of the conventional AA and AB stacking configurations, and that the expanded structural space accessible through twist engineering contains configurations with superior electrochemical properties.

The SOAP-based prediction model established in Section 2.2 provides a practical tool for extending this search. Because the model can predict the PES of untested twist angle configurations from existing DFT data with $R^2 > 0.9$, it enables rapid screening of candidate structures without requiring a full DFT-based PES calculation for each new configuration. This capability is particularly valuable for exploring tBLG structures with larger moiré periods (e.g., Σ43 and beyond), where the computational cost of exhaustive PES mapping becomes increasingly prohibitive. By combining a limited set of DFT calculations with SOAP-based predictions, the optimal twist angle for Li intercalation could be identified more efficiently across a broader range of structural candidates.

While Moirè-Ionics presents a promising frontier, several challenges remain before the potential of twist-engineered bilayer graphene electrodes can be realized in practice. First, the large-area synthesis of tBLG with precisely controlled twist angles is required. While recent advances in chemical vapor deposition (CVD) and epitaxial growth techniques have demonstrated the feasibility of producing tBLG with specific twist angles, scalable fabrication methods remain under active development. Second, the present study considers the dilute limit of a single Li atom intercalated in the moiré supercell. At higher Li concentrations relevant to practical battery operation, Li–Li interactions and collective ordering effects may modify the energy landscape and diffusion properties. Third, the long-term cycling stability and the effects of repeated intercalation/deintercalation on the structural integrity of tBLG have not been addressed and warrant further investigation.

Moreover, the adsorption energy calculations on monolayer graphene (Figure S1) reveal that the magnitude of site-dependent energy variation, quantified by the hollow/on-top/bridge energy difference on a single graphene sheet, differs substantially across elements. This quantity can serve as a simple, low-cost screening metric: elements with a large hollow/on-top/bridge energy difference are expected to show a more pronounced sensitivity of the PES to twist angle, making them promising candidates for twist-angle engineering, whereas elements with a weak site preference may show limited tunability. Because this metric requires only three single-point calculations on monolayer graphene, it provides an efficient first filter before undertaking full tBLG PES calculations. The SOAP-based prediction framework developed here could then be applied to the most promising intercalant species, enabling efficient exploration of optimal tBLG configurations beyond the Li system studied in this work.

Together, these results suggest that the combination of monolayer adsorption energetics, twist angle engineering, and machine-learning-assisted screening constitutes a general and highly transferable strategy, Moirè-Ionics, for designing next-generation two-dimensional layered materials with tailored ion transport properties.

## 3. Methodology

### 3.1 Computational Details

First-principles calculations based on density functional theory (DFT) were performed using the Vienna Ab initio Simulation Package (VASP) [38–40]. The electron–ion interactions were described within the projector augmented wave (PAW) method. The rev-vdW-DF2 [41] exchange-correlation functional was employed to account for van der Waals interactions, which has been reported to systematically reproduce the lattice constants of various layered materials [42].

The plane-wave basis set was expanded with a cutoff energy of 650 eV. Brillouin zone integration was performed using Γ-centered Monkhorst–Pack k-point meshes with a spacing of less than 0.2 $Å^{-1}$. Spin polarization was included in all calculations.

For structural optimization, the lattice parameters were fixed and only the atomic positions were relaxed. In the potential energy surface (PES) calculations, the in-plane (xy) coordinates of all atoms were fixed and relaxation was performed only along the z-axis (interlayer direction). The electronic convergence criterion was set to a total energy change of less than $1.0 \times 10^{-5}$ eV between successive self-consistent field iterations. The ionic relaxation convergence criterion was set to a maximum force of less than 0.03 eV/Å on all atoms.

### 3.2 Structure Generation

Twisted bilayer graphene (tBLG) structures were generated using the Interface Master code[43] based on coincidence site lattice (CSL) theory. CSL structures are characterized by the Σ value, and this study investigated structures ranging from Σ7 to Σ37, corresponding to twist angles from 9.43° to 32.20°. For comparison, AA stacking and AB stacking configurations of bilayer graphene were also considered.

The in-plane lattice constant of graphene was set to 2.46 Å. The initial interlayer distance for each structure was set to approximately 3.35 Å, typical of graphite, and subsequently relaxed through structural optimization. A vacuum layer of 25 Å was introduced along the z-axis to avoid spurious interactions between periodic images. The structural parameters of all generated tBLG structures are summarized in Table S1 of the Supporting Information.

### 3.3 Potential Energy Surface Calculation

To map the potential energy surface for Li intercalation, a single Li atom was systematically placed at grid points within the interlayer space of each structure. PES calculations were performed for AA stacking, AB stacking, and tBLG structures from Σ7 to Σ37.

A uniform in-plane grid was defined with a spacing of approximately 0.6–0.65 Å, and one Li atom was placed at each grid point at the midpoint of the interlayer gap as the initial position. The in-plane coordinates of all C and Li atoms were fixed, and relaxation was performed only along the z-axis.

Supercells were employed for the smaller structures (5×5 for AA and AB stacking, 2×2 for Σ7 and Σ13) to maintain a sufficiently low Li concentration and to ensure convergence of the intercalation energy with respect to cell size. The details of the PES grid for each structure are provided in Table S2.

The PES was defined relative to the most stable intercalation site within each structure:

$$PES(i,j) = E(i,j) - E_{min}$$

where $E(i,j)$ is the total energy at grid point (i, j) and $E_{min}$ is the minimum energy across all grid points.

### 3.4 Energy Analysis

#### 3.4.1 Intercalation Energy

The intercalation energy at the most stable site was calculated to evaluate the thermodynamic favorability of Li intercalation:

$$E_{int} = E_{\min}(PES) - E_{host} - E_{Li}$$

where $E_{\min}(PES)$ is the minimum total energy on the PES, $E_{host}$ is the total energy of the bilayer graphene structure without Li, and $E_{Li}$ is the total energy of an isolated Li atom.

#### 3.4.2 Diffusion Barrier

To evaluate Li diffusion kinetics, diffusion pathways were identified on the PES by tracing low-energy routes that traverse the entire periodic cell, representing the minimum energy path for sustained Li transport in a given direction. The diffusion barrier was defined as the energy difference between the maximum and minimum along this pathway:

$$E_{barrier} = \max\,[E_{path}] - \min\,[E_{path}]$$

This value represents the minimum energy required for continuous Li diffusion across the structure.

### 3.5 SOAP Descriptor and Machine Learning Model

To investigate the relationship between local atomic environment and potential energy, we employed the Smooth Overlap of Atomic Positions (SOAP) descriptor. SOAP encodes the local atomic environment around a given site as a rotationally invariant power spectrum of the neighbor density.

SOAP descriptors were computed using the DScribe package with the following parameters: cutoff radius $r_{cut}$ = 3 Å, and radial and angular expansion orders $n_{max}$ = $l_{max}$ = 3, yielding a descriptor vector of 84 dimensions for each Li position. At each PES grid point, the SOAP descriptor was evaluated for the Li atom position, capturing the arrangement of surrounding C atoms.

Ridge regression (L2-regularized linear regression) with a regularization parameter $\alpha = 0.1$ was used to map the SOAP descriptor vector to the potential energy. The model was evaluated in two settings: (1) within-structure prediction, where the model was trained and tested on different grid points of the same structure using random train/test splits, and (2) cross-structure prediction, where the model trained on one or more structures was used to predict the PES of a different structure. Prediction accuracy was assessed using the coefficient of determination ($R^2$) and mean absolute error (MAE).

## Acknowledgements

This study was supported by the Ministry of Education, Culture, Sports, Science and Technology (MEXT) (Nos. 24H00042 and 26H000), and New Energy and Industrial Technology Development Organization (NEDO). Some computations were carried out using the computer resource offered by Research Institute for Information Technology, Kyushu University.

## Data Availability Statement

The simulated data and PES and constructed model in this study are available from the corresponding author upon reasonable request.

**Supporting Information**

**Twist-Angle Engineering of Moiré Potentials for High-Performance Ionics in Bilayer Graphene**

Gen Fukuzawa, Yebin Lee, and Teruyasu Mizoguchi

## S1 Interlayer Binding Energy

The interlayer binding energy was calculated to evaluate the thermodynamic stability of each bilayer graphene structure:

$$E_{coh} = (E_{bilayer} - 2 \times E_{monolayer})/N$$

where $E_{bilayer}$ is the total energy of the bilayer graphene, $E_{monolayer}$ is the total energy of the monolayer graphene, and N is the total number of atoms in the bilayer. Normalization per atom enables comparison across structures of different sizes.

## S2. Structural Parameters

| Structure | (N, M) | Twist angle (°) | Interlayer distance (Å) | Binding energy (meV/atom) |
|---|---|---|---|---|
| AA | - | 0 | 3.52 | −22.65 |
| AB | - | 0 | 3.35 | −26.90 |
| Σ7 | (1, 2) | 21.79 | 3.41 | −24.47 |
| Σ13 | (1, 3) | 32.20 | 3.41 | −26.44 |
| Σ19 | (2, 3) | 13.17 | 3.42 | −26.17 |
| Σ31 | (1, 5) | 17.90 | 3.41 | −24.66 |
| Σ37 | (3, 4) | 9.43 | 3.41 | −24.74 |

**Table S1.** Structural parameters and interlayer binding energies of the bilayer graphene structures investigated in this study.

## S3. PES Calculation Details

| Structure | Supercell | Number of C atoms | Computational cell size (Å²) | Grid spacing (Å) | Number of divisions | Number of grid points |
|---|---|---|---|---|---|---|
| AA | 5×5 | 100 | 151.75 | 0.62 | 20 | 400 |
| AB | 5×5 | 100 | 151.75 | 0.62 | 20 | 400 |
| Σ7 | 2×2 | 112 | 169.96 | 0.65 | 20 | 400 |
| Σ13 | 2×2 | 208 | 315.64 | 0.63 | 28 | 784 |
| Σ19 | 1×1 | 76 | 115.33 | 0.60 | 18 | 324 |
| Σ31 | 1×1 | 124 | 188.18 | 0.65 | 21 | 441 |
| Σ37 | 1×1 | 148 | 224.60 | 0.62 | 24 | 576 |

**Table S2.** Details of the PES calculations for each structure.

## S4. Adsorption Energies on Monolayer Graphene

To understand the origin of the PES patterns observed in bilayer graphene, we computed the adsorption energies of single metal atoms (up to period 6, excluding lanthanoids) on monolayer graphene at three representative sites: the hollow site (center of the hexagonal ring), the bridge site (midpoint of a C–C bond), and the on-top site (directly above a C atom). The adsorption energy is defined as:

$$E_{ads} = E_{sys} - E_{sheet} - E_{atom}$$

where $E_{sys}$ is the total energy of the adsorbed system, $E_{sheet}$ is the total energy of the graphene sheet, and $E_{atom}$ is the total energy of the isolated atom.

Figure S1 shows the computed adsorption energies across the periodic table. For Li, the hollow site is clearly the most stable (−1.25 eV), while the bridge (−0.96 eV) and on-top (−0.95 eV) sites are nearly equivalent. This indicates that the primary energetic distinction is between hollow and non-hollow sites, which underlies the PES patterns discussed in Section 2.1.1 of the main text.

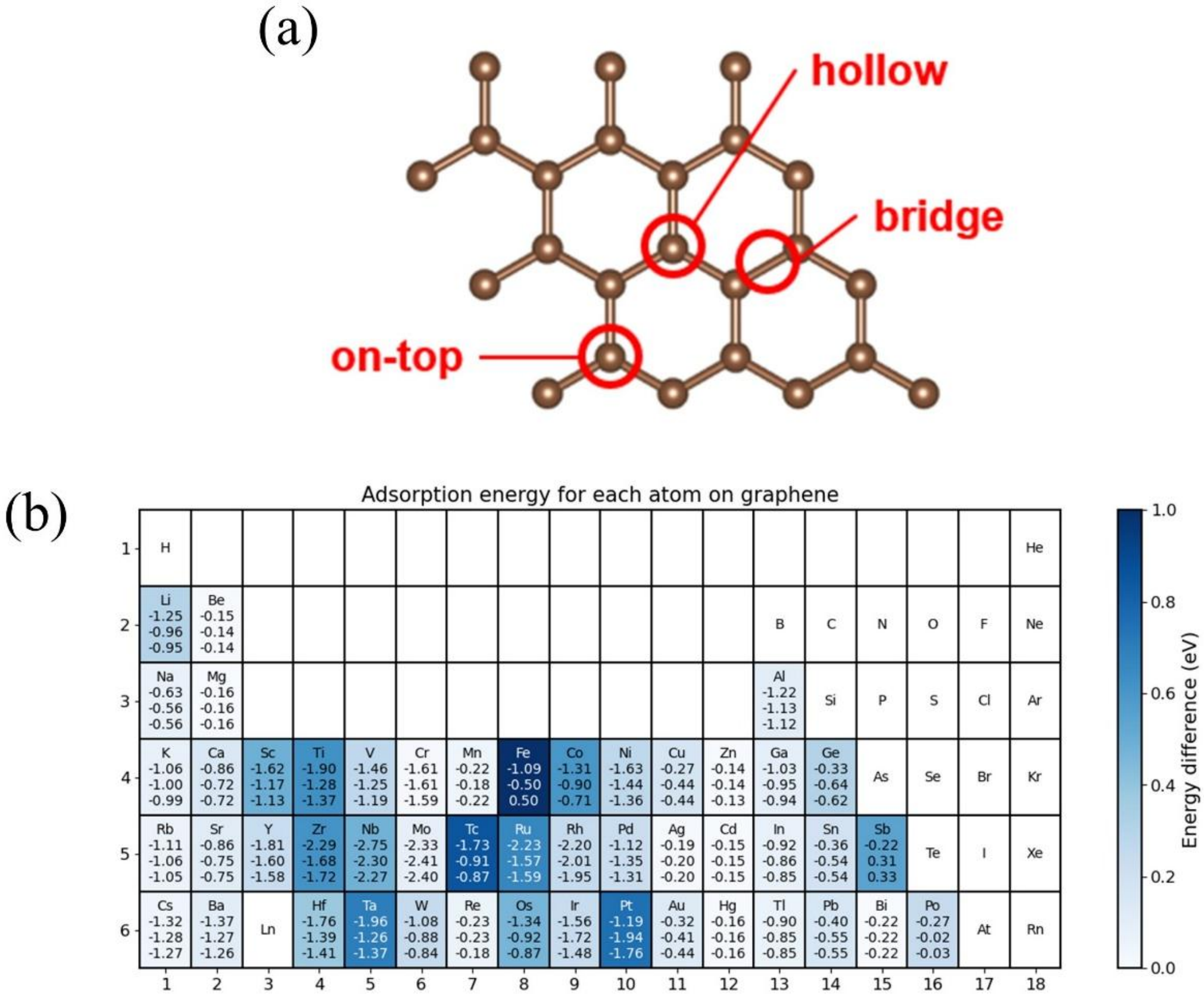

**Figure S1.** (a) Adsorption sites on monolayer graphene: hollow, bridge, and on-top. (b) Adsorption energies of single metal atoms on monolayer graphene, displayed in periodic table format. For each element, the three values correspond to (top) hollow, (middle) bridge, and (bottom) on-top sites.

**S5. PES Maps and Diffusion Pathways**

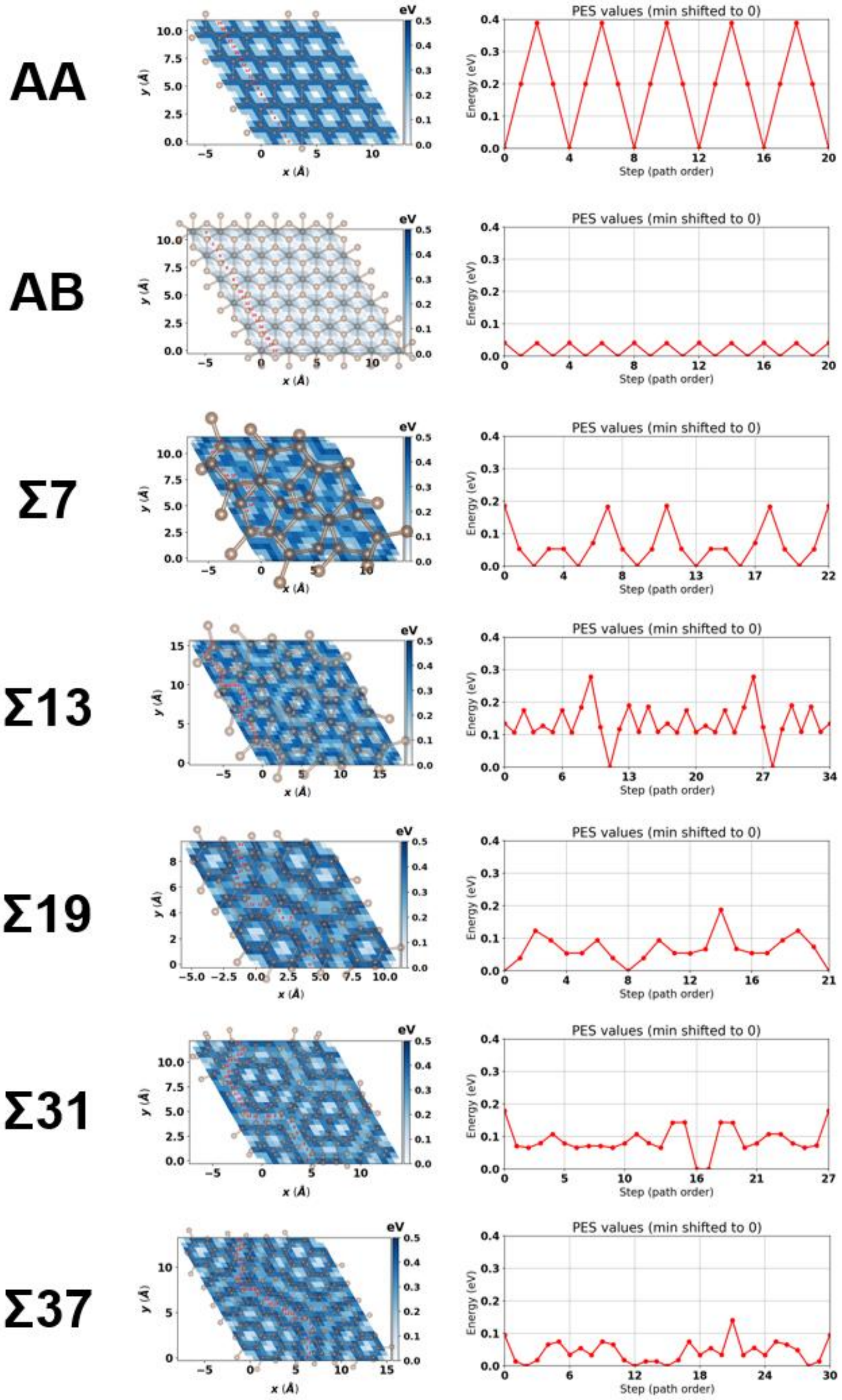

**Figure S2.** Potential energy surface maps (left) and energy profiles along the selected diffusion pathways (right) for all structures. Red dashed lines and numbered points on the PES maps indicate the selected pathways and their step order. The energy profiles show the potential energy at each step along the pathway, with the minimum shifted to zero. The diffusion barrier corresponds to the maximum value in each profile.